\documentstyle[aps,prb,multicol]{revtex}
\begin{document}

\include{psfig}

\draft \title{\bf Field dependence of the microwave resistivity in 
SmBa$_{2}$Cu$_{3}$O$_{7}$ thin films.}

\author{E. Silva$^{(1)}$\footnote[2]{To whom correspondence should be 
addressed.  E-mail: silva@fis.uniroma3.it}, N. Pompeo$^{(1)}$, L. 
Muzzi$^{(1)}$\footnote[3]{Present Address: ENEA, Frascati, Roma, Italy.}, 
R. Marcon$^{(1)}$, S. Sarti$^{(2)}$, M. Boffa$^{(3)}$, A. M. Cucolo$^{(3)}$}

\address{$^{(1)}$Dipartimento  di Fisica "E. Amaldi" and Unit\`a INFM, 
Universit\`a di Roma Tre, Via della Vasca Navale 84, 00146 Roma, Italy\\
$^{(2)}$ Dipartimento di Fisica and 
Unit\`a INFM, Universit\`a "La Sapienza", 00185 Roma, Italy\\
$^{(3)}$ Dipartimento  di Fisica and Unit\`a INFM, Universit\`a di 
Salerno, Baronissi, Salerno, Italy}

\date{preprint}

\maketitle

\begin{abstract}
We report measurements of the microwave complex resistivity at 48 GHz 
in SmBa$_{2}$Cu$_{3}$O$_{7-\delta}$ thin films.  Measurements are 
performed with a moderate magnetic field, $\mu_{0}H<$ 0.8T, applied 
along the $c$-axis.  We find that the complex resistivity presents 
clear sublinear field dependences, and that the imaginary part is 
remarkably sensitive to the moderate magnetic field.  Interpretation 
considering an unusually strong pinning leads to very anomalous field 
dependences of the single-vortex viscosity and of the pinning 
constant.  By contrast, allowing for a significant effect of the 
magnetic field on the depletion of the condensate, the data are 
quantitatively described by the simple free-flow-like expression, 
supplemented with two-fluid conductivity.  In this frame, we obtain 
the vortex viscosity from the data.  We compare vortex viscosity in 
SmBa$_{2}$Cu$_{3}$O$_{7-\delta}$ and in 
YBa$_{2}$Cu$_{3}$O$_{7-\delta}$.

\end{abstract}


\begin{multicols}{2}

\section{Introduction}
\label{intro}
The microwave response is a source of important information in 
superconductors.  In high-$T_{c}$ superconductors (HTCS) various 
microwave techniques have been used to get information, among the 
others, on the symmetry of the order parameter, on the vortex 
parameters such as the vortex viscosity and pinning frequency 
\cite{golos,tsuchiya,owliaei} and on the temperature dependence of the 
superfluid fraction (via the measurement of the temperature dependence 
of the London penetration depth \cite{hardy}).  In short, microwave 
measurements are a very powerful tool to study the superconducting 
state.\\
While YBa$_{2}$Cu$_{3}$O$_{7-\delta}$ (YBCO) and, to a minor extent, 
Bi$_{2}$Sr$_{2}$CaCu$_{2}$O$_{8+x}$ (BSCCO) have been the subject of 
intensive experimental investigation, other HTCS did not receive the 
same attention.  In particular, a very few reports dealt with the 
vortex-state microwave response in rare-earth substituted 123 
compounds, mainly on Dy \cite{bhangale} and Gd-substituted 
materials.\cite{blackstead,silvaGd} In fact, RE-substituted 123 
compounds present potentially interesting features in the vortex 
state: in particular, they often give enhanced irreversibility lines 
\cite{irrline} and in some case a $T_{c}$ slightly higher than in 
YBCO. To what extent those features can change the microwave 
properties in the vortex state is not known, nor it is clear what 
effect substitutions can have on the pinning at high frequencies.  Aim 
of this paper is to present an extensive study of the microwave 
response of SmBa$_{2}$Cu$_{3}$O$_{7-\delta}$ (SmBCO) in the vortex 
state in moderate fields, up to the critical temperature.  It will be 
shown that the complex resistivity shows up an unexpectedly strong 
field dependence.  In particular, the imaginary part increases 
noticeably with the applied magnetic field at low temperatures, 
becomes nearly field-independent at a sample-dependent temperature 
around $T/T_{c}\sim$ 0.9 and then decreases with the application of 
the field up to $T_{c}$.  In addition, both the real and imaginary 
parts of the resistivity present clear sublinear field dependences.  
We show that all those findings cannot be easily understood within the 
conventional framework for vortex motion at high frequencies.  By 
contrast we find that, by assuming a significant effect of the 
magnetic field on the superfluid depletion, all the temperature and 
field dependences are well described by the simple viscous vortex 
motion supplemented with the two-fluid conductivity.  We obtain almost 
the same temperature-dependent vortex viscosity in two different 
samples, in agreement with its intrinsic origin. Consistently, the temperature 
dependence of the vortex viscosity in SmBCO is identical to the one 
measured in YBCO.\\

\section{Samples preparation and experimental results}
\label{exp}

The SmBa$_{2}$Cu$_{3}$O$_{7-\delta}$ films were deposited on (100)K 
LaAlO$_{3}$ substrates by a high oxygen pressure sputtering technique.  
Even if these substrates show higher mismatch in comparison with the 
SrTiO$_{3}$, the use of LaAlO$_{3}$ is mandatory for microwave studies 
and applications due to its low loss tangent.  Sintered 
targets\cite{hitec} of 35 mm diameter and of 2 mm thickness were used.  
The nominal composition of the metal elements (Sm:Ba:Cu) in the 
targets was 1:2:3.  All the films were deposited at high temperatures, 
with the substrates placed onto an inconel heater whose temperature 
was monitored by a thermocouple embedded into the furnace.  The films 
were grown in a vacuum chamber at constant oxygen pressure $P_{O_{2}}$ 
and substrate temperature $T_{d}$ during the deposition.  The 
tetragonal to orthorhombic phase transition was accomplished after the 
deposition process by cooling the samples in an oxygen pressure of one 
atmosphere from $T_{d}$ to the annealing temperature, $T_{a}$, kept 
constant for the time of the annealing process, $t_{a}$.  The samples 
were finally cooled down to the room temperature.  The main deposition 
parameters are: $T_{d}=$ 910$^{\mathrm{\circ}}$C, $P_{O_{2}}=$ 2.6 
mbars, $T_{a}=$ 560$^{\mathrm{\circ}}$C, $t_{a}=$ 15 min.  The crystal 
structure of the films was examined by X-ray $\theta -2\theta$ 
diffraction.  The X-ray diffraction pattern showed that the growth of 
the film onto a LaAlO$_{3}$ substrate was single phased with 
preferential orientation along the $c$ axis.  The rocking curve of the 
(005) peak exhibits a typical full width at half maximum (FWHM) of 
less than 0.2$^{\mathrm{\circ}}$.  The surface morphology of the films 
was observed by atomic force microscopy (AFM) at room temperature.  
2D-nucleation growth was observed, and roughness was found to be 
$\sim$ 2 nm on 1 $\mathrm{\mu}$m $\times$ 1 $\mathrm{\mu}$m area.  
Film thickness was typically $d\approx$ 200 nm.  Two films were 
prepared, with $T_{c}\simeq$ 86.5 K, as determined from the inflection 
point of the temperature-dependent real resistivity.  The dc 
resisistivity at 100 K was $\rho$(100)$\simeq$300$\mu\Omega$cm in both 
samples.  Further details on sample preparation and structural 
characterization have been reported elsewhere.\cite{cucolo}\\
The microwave response was measured by the end-wall cavity technique 
at 48.2 GHz with a moderate magnetic field applied along the $c$-axis.  
The maximum attainable field was $B\simeq \mu_{0}H\leq$0.8 T. $Q$ 
factor and resonance frequency $\nu_{0}$ were measured as a function 
of the magnetic field at various fixed temperatures in the range 60-90 
K. From the field induced change of $Q$ and $\nu_{0}$ the field 
induced change of the effective surface impedance $\Delta Z_{s}^{eff} 
= Z_{s}^{eff}(H,T)-Z_{s}^{eff}(0,T)$ could be obtained.\cite{silvaMST} 
Due to the small thickness of the film, the thin film approximation 
applies to the measurements,\cite{silvaSUSTthin} so that $\Delta 
Z_{s}^{eff}=\Delta\tilde\rho/d=\Delta\rho_{1}/d+{\mathrm{i}} 
\Delta\rho_{2}/d$, where $\tilde\rho$ is the complex resistivity.  It 
must be stressed that the absolute value of $\tilde\rho$ is affected 
by errors in the evaluation of the geometrical factor and of the film 
thickness, and in the calibration of the 
cavity.\cite{ceremuga,silvareview} The first two are eliminated and 
the third strongly reduced by working with the reduced complex 
resistivity change, $\Delta \tilde r=\Delta r_{1}+{\mathrm{i}}\Delta 
r_{2}=\Delta\rho_{1}/\rho_{0}+\mathrm{i} \Delta\rho_{2}/\rho_{0}$, 
with $\rho_{0}=\rho$(100 K).  In the following we will work with the 
reduced quantity, since (as it will be clear later) the absolute 
values will be useful only in the discussion of the details of the 
parameters.\\
We now present the main features of our data.  Two samples were 
investigated, and both presented the same experimental behaviour.  In 
the following, we focus on measurements taken more 
systematically in sample A. In Figures \ref{r1vsB} and \ref{r2vsB} we 
report the field-sweeps at several temperatures for the real and 
imaginary parts.  The real part, $\Delta r_{1}$, increases with the field, 
with increasing amplitude up to a typical temperature $T_{max}\approx$ 
82 K (Figure \ref{r1vsB}a), above which the application of the field 
has a lower and lower effect by increasing the temperature (Figure 
\ref{r1vsB}b).  The imaginary part, $\Delta r_{2}$, increases with the field 
at lower temperatures, becomes approximately field independent at 
$T_{0}\approx$ 80 K and then starts to decrease with the field (Figure 
\ref{r2vsB}) above $T_{0}$. The same features are observed in sample 
B, with $T_{0}\approx$ 77 K.  Both $\Delta r_{1}$ and $\Delta r_{2}$ 
exhibit a sublinear behaviour with respect to the applied field.  The 
explicit field dependences of $\Delta r_{1}$ and $\Delta r_{2}$ can be 
directly identified by replotting the same data of Figures \ref{r1vsB} 
and \ref{r2vsB} vs $\sqrt{B}$ (Figures \ref{r1vssqrtB} and 
\ref{r2vssqrtB}).  It is immediately seen that $\Delta 
r_{2}\propto\sqrt{B}$, while $\Delta r_{1}$ has an upward curvature.  
Within the experimental uncertainty, $\Delta 
r_{1}$ is the sum of a linear and a square-root term in the applied 
field.  Summarizing, the full body of our data in SmBCO can be 
described by the empirical relations:
\begin{equation}
\label{R1}
\Delta r_{1}=b_{1}(T) B+a_{1}(T)\sqrt{B}
\end{equation}
\begin{equation}
\label{R2}
\Delta r_{2}=a_{2}(T)\sqrt{B}
\end{equation}
Interestingly, these field dependences do not vary, neither when 
$\Delta r_{2}$ changes from positive to negative, nor when the 
temperature raises so that the measurements cross the irreversibility 
line (by increasing the temperature from 65 K to $T_{c}$ we certainly 
cross $B_{irr}$), nor when the temperature is raised above $T_{max}$.  
Only very close to $T_{c}$ (approximately 1 K below, where strong 
fluctuational effects become predominant) the field dependences 
empirically found appear to slightly modify and the empirical 
equations above no longer describe the data very accurately.  All the 
obtained coefficients in the two samples studied have the same 
behavior as a function of the temperature, as reported in Figure 
\ref{a1a2} for $a_{1}$ and $a_{2}$ and in Figure \ref{b1} for $b_{1}$.  
This is a clear indication (a) against the relevance of pinning at our 
measuring frequency, and (b) in favour of some ``intrinsic'' origin 
for the behaviour of the resistivity in the mixed state.\\
Given the very clear field dependences of the complex resistivity, we 
search for a theoretical frame able to describe our data, taking into 
account that no features appear when crossing 
the irreversibility line.  In the following Section we briefly 
summarize the reference frame that we adopt for the theoretical 
description of the microwave response in the vortex state.

\section{Theoretical summary}
\label{th}

In principle, the response of a superconductor in the mixed state is 
dictated by the motion of vortices and of charge carriers.  The latter 
originates from superfluid as well as normal carriers, and 
field-induced changes of the quasiparticle (QP) density have a direct 
effect on both the real and the imaginary part.  A general frame that includes 
all those contributions is the Coffey-Clem (CC) theory \cite{cc} (a 
very similar model has been independently derived by Brandt 
\cite{brandt}), based on the assumption of the two-fluid conductivity 
and of a periodic vortex pinning potential.\\
The general CC expression, including vortex, superfluid and QP 
contributions, is rather complex.  In particular, the combined effect 
of pinning and creep make even the vortex motion alone of difficult 
description.  In order to clarify the role of pinning, we first focus 
the attention below the irreversibility line where its effect is 
maximum.  In the latter case, the characteristic frequency for the 
vortex diffusion reduces to the pinning frequency 
$2\pi\nu_{p}=k_{p}/\eta$, where $\eta$ is the vortex viscosity and 
$k_{p}$ the pinning constant (Labusch parameter).  The resulting 
expression for the microwave complex resistivity 
$\tilde\rho=\rho_{1}+\mathrm{i} \rho_{2}$ in the vortex state can be 
then cast in the form:
\begin{equation}
\label{cc1}
\rho_{1}= \frac{1}{1+(\frac{\sigma_{R}}{\sigma_{I}})^{2}} 
\left[\rho_{ff}\frac{1+\frac{\nu_{p}}{\nu}\frac{\sigma_{R}}{\sigma_{I}}}{1+(\frac{\nu_{p}}{\nu})^{2}} 
+\frac{1}{\sigma_{R}}\left(\frac{\sigma_{R}}{\sigma_{I}}\right)^{2}\right]
\end{equation}
and
\begin{equation}
\label{cc2}
\rho_{2}= \frac{1}{1+(\frac{\sigma_{R}}{\sigma_{I}})^{2}} 
\left[\rho_{ff}\frac{\frac{\nu_{p}}{\nu}-\frac{\sigma_{R}}{\sigma_{I}}}{1+(\frac{\nu_{p}}{\nu})^{2}} 
+\frac{1}{\sigma_{I}}\right]
\end{equation}
Where $\nu$ is the measuring frequency, $\rho_{ff}=\Phi_{0}B/\eta$ is 
the so-called flux- resistivity, and 
$\sigma_{R}-{\mathrm{i}}\sigma_{I}$ is the complex conductivity in 
absence of motion of vortices.  The latter is temperature, field and 
frequency dependent, and contains the contributions of superfluid and 
QP, through their fractional densities and the QP scattering time.\\
These expressions have been sometimes used for the analysis of the 
microwave response in the vortex state.\cite{golos} However, in most 
cases it was judged that the field-dependence of the QP and superfluid 
densities was negligible, and that at low enough temperatures 
$\sigma_{I}\gg\sigma_{R}$.  Within these approximations, the 
Coffey-Clem (CC) prediction for the field-induced changes to the 
resistivity reduces to the well-known Gittleman-Rosenblum (GR) 
model \cite{gittle} (in the GR model the QP and superfluid 
contributions are not considered, so the response is described by the 
vortex motion alone):
\begin{equation}
\label{gr1}
\Delta \rho_{1}(B)=\rho_{1}(B)-\rho_{1}(B=0)=  
\frac{1}{1+(\frac{\nu_{p}}{\nu})^{2}}\rho_{ff}
\end{equation}
and
\begin{equation}
\label{gr2}
\Delta \rho_{2}(B)=\rho_{2}(B)-\rho_{2}(B=0)= 
\frac{\frac{\nu_{p}}{\nu}}{1+(\frac{\nu_{p}}{\nu})^{2}}\rho_{ff}
\end{equation}
The latter model was in fact extensively used for the analysis of the 
microwave resistivity or surface impedance (see Ref.\cite{golos} and 
references therein).  When it is possible to apply this conventional 
analysis, the vortex parameters $\nu_{p}$ and $\rho_{ff}$ can be 
easily obtained by inverting the data through Equations 
\ref{gr1},\ref{gr2}.  The vortex viscosity is calculated as 
$\eta=\Phi_{0}B/\rho_{ff}$.  Finally, the pinning constant comes from the 
definition of $\nu_{p}$.\\
In the following Section, we discuss first our data below the 
irreversibility line in terms of the GR model, and then within the CC 
model allowing for the inclusion of a possibly relevant effect of the 
magnetic field on the QP and superfluid fractional densities.  The 
emerging frame strongly points toward the latter interpretation.

\section{Discussion}
\label{disc}

When discussing our data on SmBCO we are faced with two main 
experimental features: the field dependences of $r_{1}$ and $r_{2}$, 
which are clearly sublinear, and the very relevant increase of the 
imaginary part in even moderate fields.  Since the role of pinning 
reveals itself mostly on the imaginary part, one might be tempted to 
assign the strong field dependence of $\Delta r_{2}$ simply to a 
strong pinning.  Within this quite conventional view, one would apply 
the GR model and, following the procedure explained in Section 
\ref{th}, would obtain $\nu_{p}$, $\eta$, $k_{p}$ directly from the data.  
The result of the procedure is reported in Figure \ref{conventional} 
for several temperatures.  As can be seen, the so-obtained pinning 
frequency would be a very weak function of the applied field, in 
agreement with commonly reported dependences.\cite{golos}  However, 
the viscosity $\eta$ would present a clear increase with the field 
(approximately as $\sim\sqrt{B}$), as reported in Figure 
\ref{conventional} (panel b).  This dependence is not easily 
explained.  In fact, a similar behaviour at high fields has been 
tentatively explained in Bi$_{2}$Sr$_{2}$CuO$_{6}$ 
(Ref.\cite{matsuda}) in terms of a peculiar field dependence of the 
quasiparticle relaxation time in a d-wave superconductor, appearing 
when the intervortex distance becomes smaller than the mean free path.  
However, the model would explicitly predict the usual {\it field 
independence} at low fields.  Due to our field and temperature ranges 
(by far lower than the temperature-dependent upper critical field), 
this picture does not seem very convincing.  The anomalies become even 
more evident when, within the same GR model, we try to obtain the 
pinning constant $k_{p}$ from the vortex pinning frequency and the 
vortex viscosity: as reported in Figure \ref{conventional} (panel c), 
the so-obtained pinning constant would {\it increase} with the field 
(again, approximately as a square root).  This behaviour does not seem 
reasonable.  In particular, we note that in Ref.\cite{matsuda} $k_{p}$ 
was found to be constant at low fields, and to decrease at higher 
fields, as expected in high-frequency measurements.  We conclude that 
an explanation of our data in terms of the GR model is at least 
questionable.\\
We now show that the empirical field dependences for the complex 
resitivity can be immediately derived within a CC-like model, by 
including a field dependence of the QP fractional density in agreement 
with the existence of lines of nodes in the superconducting gap.\\
We first critically discuss the assumptions underlying the CC model.  
The first assumption, the applicability of the two-fluid model, is 
reasonably justified by comparison with microwave 
studies\cite{bonnPRL92,bonnPRB93} of YBCO, which is the 
superconducting cuprate most similar to SmBCO. In particular, the 
imaginary part of the conductivity can be written as: 
$\sigma_{I}(T,B=0)=\frac{x_{s0}(T)}{2\pi\nu\mu_{0}\lambda_{0}^{2}}\equiv\sigma_{I0}x_{s0}(T)$, 
where $\lambda_{0}$ is the zero-temperature London penetration depth, 
and $x_{s0}$ is the temperature-dependent superfluid fraction, which 
is found\cite{bonnPRB93} to be $x_{s0}\propto 1-t^{2}$, with 
$t=T/T_{c}$, to a good approximation in the full temperature range 
from low temperatures up to very close to $T_{c}$.  Moreover, it was 
found that the normal fluid fractional density $x_{n0}$ decreased with 
decreasing temperature below $T_{c}$.  In addition, no significant 
frequency dependence was observed \cite{hosseini} in the real 
conductivity up to 75 GHz for temperatures $T>$ 60 K indicating that 
at 50 GHz and above $T/T_{c}>\frac{2}{3}$ the real part of the 
conductivity can be written as $\sigma_{R}(T,B=0)=\frac{ne^{2}\tau}{m} 
x_{n0}(T)\equiv\sigma_{R0}x_{n0}(T)$, where $\tau$ is the 
quasiparticle relaxation time.  \\
Regarding the chosen description of the vortex motion, the assumption 
of periodic pinning potential is believed not to strongly affect the 
main features of the response, until the probing frequency is high 
enough to displace the vortices a small fraction of the intervortex 
distance, as is the case at high microwave frequencies.  In fact, 
swept-frequency Corbino disk measurements in the frequency range 6-20 
GHz\cite{sartiCM} have shown that the frequency dependence of 
$\Delta\rho_{1}$ was well described by the Coffey-Clem expression for 
vortex dissipation, and gave estimates of the pinning frequency 
$(\nu_{p}/\nu)^{2}\sim$ 5$\times$10$^{-2}$ at 50 GHz and above 60 
K.\cite{sartiEU}\\
Assuming that the main features of YBCO apply also to the closely 
related compound SmBCO, we compare our data to a CC-like model where 
the field-induced superfluid depletion is explicitly included, and we 
assume $\nu_{p}/\nu\ll 1$. The latter assumption is further supported 
by the experimental fact that our data do not 
show any detectable feature with increasing temperature from below to 
above the irreversibility line.  In addition, at small enough 
fields, not too close to $T_{c}$ we have $\rho_{ff} \sigma_{R}\ll 1$ 
and $\frac{\sigma_{R}}{\sigma_{I}}<$ 1, so that Eq.s 
\ref{cc1},\ref{cc2} simplify considerably and we 
obtain the approximate expressions:
\begin{equation}
\label{cc1appr}
\rho_{1}\simeq 
\frac{\rho_{ff}}{1+(\frac{\sigma_{R}}{\sigma_{I}})^{2}}+
\frac{1}{\sigma_{R}}\frac{(\frac{\sigma_{R}}{\sigma_{I}})^{2}}{1+(\frac{\sigma_{R}}{\sigma_{I}})^{2}}
\end{equation}
and
\begin{equation}
\label{cc2appr}
\rho_{2}\simeq
\frac{1}{\sigma_{I}}\frac{1}{1+(\frac{\sigma_{R}}{\sigma_{I}})^{2}}
\end{equation}
It can also be shown that, within the approximations described above, 
Eq.s \ref{cc1appr},\ref{cc2appr} remain valid above and below the 
irreversibility line: these approximate equations represent (a) the 
microwave response as given by free oscillations of flux lines 
(analytically described by the flux flow resisitivity), given by the 
first term in Eq.\ref{cc1appr}, and (b) the contribution from 
superfluid and QP conductivity, as given by Eq.\ref{cc2appr} and by 
the second term in Eq.\ref{cc1appr}.\\
An explicit field dependence of the two-fluid conductivity has to be 
inserted in these equations in order to compare them to the data.  In 
a clean, conventional s-wave superconductor the superfluid density in 
the mixed state would decrease, and the QP fraction would increase, 
due to the density of states in the vortex cores.  The effect would be 
proportional to the number of vortices, thus linear in $B$.  However, 
it has been shown that in a superconductor with lines of nodes in the 
superconducting gap delocalized states (outside of the vortex cores) 
give a very significant contribution, exhibiting finite density of 
states at the Fermi level.\cite{volovik,won,yip} The DOS is 
proportional to the number of vortices and to the spacing between 
them, thus leading to $x_{n}\sim \sqrt{B}$ increase of the QP 
fractional density.  This effect has been experimentally observed in 
the specific heat of YBCO,\cite{moler} and in measurements of the QP 
fractional increase with sub-TeraHertz spctroscopy in BSCCO thin 
films.\cite{mallozzi}\\
We then incorporate the field dependence in the two-fluid conductivity 
through the superfluid and QP fractional densities\cite{notatau} as 
$x_{s}(T,B)=x_{s0}(T) (1-b^{\alpha})$ and 
$x_{n}(T,B)=1-\left(1-x_{n0}(T)\right) (1-b^{\alpha})$, with 
$b=B/cB_{c2}(T)$, $c\sim o(1)$ (Ref.\cite{won}) and $\alpha=\frac{1}{2}$ as 
appropriate for a superconductor with lines of nodes in the 
superconducting gap.\cite{mallozzi,volovik,won}\\
Interestingly, taking into account that $b^{\alpha}$ is a small 
parameter, one can expand the two-fluid terms in Eq.s \ref{cc1appr} 
and \ref{cc2appr}. One finally gets for 
the field variations of the normalized resistivity:
\begin{eqnarray}
\nonumber
\Delta r_{1}(B)\simeq \frac{1}{1+S^{2}}\frac{\Phi_{0}}{\eta 
\rho_{0}}B+\frac{1}{\sigma_{R0}\rho_{0}}\frac{1}{x_{n0}}
\frac{S^{2}}{\left(1+S^{2}\right)^{2}}\times\\
\nonumber
\times\left[1+\frac{x_{s0}}{x_{n0}}+S^{2}\left(1-\frac{x_{s0}}{x_{n0}}\right)\right] 
\left(\frac{B}{cB_{c2}}\right)^{\frac{1}{2}}\\
\label{cc1final}
\equiv b_{1}(T) B + 
a_{1}(T) B^{\frac{1}{2}}
\end{eqnarray}
and
\begin{eqnarray}
\nonumber
\Delta r_{2}(B)\simeq\\
\nonumber
\frac{1}{\sigma_{R0}\rho_{0}} \frac{1}{x_{n0}}
\frac{S}{\left(1+S^{2}\right)^{2}} 
\left[1-S^{2}\left(1+2\frac{x_{s0}}{x_{n0}}\right)\right] 
\left(\frac{B}{cB_{c2}}\right)^{\frac{1}{2}}\\
\label{cc2final}
\equiv a_{2}(T) 
B^{\frac{1}{2}}
\end{eqnarray}
where we have defined 
$S=\frac{\sigma_{R0}}{\sigma_{I0}}\frac{x_{n0}}{x_{s0}}\equiv 
S_{0}\frac{x_{n0}}{x_{s0}}$.  These two equations give exactly the 
functional field dependences that have been empirically found in the 
data.  This is one of the main results of this paper: the experimental 
field dependence of the complex resistivity, not explainable in the 
framework of fluxon motion alone, is recovered by a model that 
includes simple free flux line oscillation (corresponding to flux flow 
resistivity) and the essential role of superfluid depletion in a 
superconductor with lines of nodes in the gap.  We note that the 
inclusion of a vortex motion term in the imaginary resistivity would 
have given an additional {\it linear} term in Eq.\ref{cc2final}.  To a 
great accuracy, no $B$-linear term is observed in the data, giving 
further consistency to the approximations used to derive Eq.s 
\ref{cc1final},\ref{cc2final}.  As a further relevant result, we note 
that the $a_{1}$ and $a_{2}$ coefficients here defined present the 
same qualitative features as in the experimental values: on one side, 
the study of Eq.\ref{cc1final} reveals that $a_{1}$ can present a 
peak; more important, it is immediately seen that $a_{2}$ undergoes a 
sign change at a temperature such that: %
\begin{equation} 
S_{0}=\left[\frac{x_{n0}}{x_{s0}}\left(2+\frac{x_{n0}}{x_{s0}}\right)\right]^{-\frac{1}{2}}
\label{vincolo}
\end {equation}
Summarizing, the model here developed contains all the 
experimental features present in the data: the field dependences of 
the complex resistivity and the behavior with the temperature of $a_{1}$ and $a_{2}$.\\
For a quantitative fit of the data with the theoretical expressions, 
one has to determine the temperature dependences of the parameters 
contained in Eq.s \ref{cc1final},\ref{cc2final}.  Exploiting the 
similarities with YBCO, and consistently with the indication of the 
existence of lines of nodes in the gap, we have taken 
$x_{s0}=\left(1-t^{2}\right)$ and $x_{n0}=t^{2}$, and 
$B_{c2}=B_{c20}\left(1-t^{2}\right)$.  To our knowledge there are no 
detailed studies of the finite frequency conductivity in SmBCO that 
could give indications on the temperature dependence of the QP 
scattering time: we then make the very crude assumption that the QP 
scattering time does not change much from above to below $T_{c}$.  
With this choice, one can write (apart small corrections due to the 
temperature dependence of the normal state resistivity, that we 
neglect here) $\sigma_{r0}\rho_{0}\simeq$1, and $S_{0}\simeq const$.  
With this choice, on the basis of the constraint given by 
Eq.\ref{vincolo} we have $S_{0}=$0.146 and 0.212, in sample A and B, 
respectively.\cite{notafit} The simultaneous fits of $a_{1}(T)$ and 
$a_{2}(T)$ contain only $cB_{c20}$ as a common scale factor.  In 
Figure \ref{a1a2} we plot the coefficients $a_{1}$ and $a_{2}$, 
compared with the theoretical curves computed on the basis of Eq.s 
\ref{cc1final},\ref{cc2final} with $cB_{c20}=$330 T and 230 T as scale 
factors for sample A and B, respectively.  The numerical values of 
$S_{0}$ are larger by a factor $\sim$ 30 than expected with 
$\rho_{0}\simeq$ 280 $\mu\Omega$cm and $\lambda_{0}\sim$ 2000\AA. 
Neglected effects, such that a temperature dependent quasiparticle 
scattering time,\cite{bonnPRB93} a zero-temperature superfluid fraction less than 
unity, a two-component order parameter\cite{srikanth} might be 
responsible for this enhancement.  Nevertheless, given the crudeness 
of the model and the absence of fit parameters, the theoretical 
expectations describe with surprising accuracy our data.  We believe 
that the essential physics of the field dependent microwave response 
is related to the effects extensively described in this paper.\\
Finally, we discuss the fluxon dynamics.  Within the present 
interpretation, it is entirely described by the coefficient $b_{1}$.  
Using for $S=S_{0}\frac{x_{n0}}{x_{s0}}$ the determinations obtained 
from $a_{1}$ and $a_{2}$ above, we immediately get the fluxon 
viscosity $\eta$, as reported in Figure \ref{b1}. As can be seen, the 
data for $\eta$ attain the same value in both samples, indicating that 
the physics in the vortex core is related to sample-independent 
processes.\\
Further insights can be gained by examining the data within the 
Bardeen-Stephen model for the viscosity:\cite{bs} in this case 
$\eta=\Phi_{0}B_{c2}/\rho_{n}$. Since (within 10\%) $\rho_{n}$ is the 
same in both samples, we deduce that also the upper critical field is 
the same.  From that, we get for the constant $c$ the relation 
$c_{A}/c_{B}\simeq$1.44 (subscripts refer to the sample).  Comparison 
with vortex viscosity data taken in YBCO\cite{silvaVORTEX} shows that: 
(a) the temperature dependence is the same in SmBCO and YBCO (minor 
differences appear only close to the critical temperature), and (b) 
the data of YBCO scale with a factor of 2 on the corresponding data in 
SmBCO, as we show in Figure \ref{etainsieme}.  Since the ratio of 
the normal state resistivities at 100 K in the YBCO sample here 
reported and in the SmBCO samples is 
$\rho_{0}$(SmBCO)/$\rho_{0}$(YBCO)$\simeq$ 3, and 
$B_{c20}$(YBCO)$\simeq$ 165 T,\cite{silvaSUST} using again the 
Bardeen-Stephen model we obtain $B_{c20}$(SmBCO)$\simeq$ 250 T, 
$c_{A}\simeq$1.32, $c_{B}\simeq$0.92, in agreement with the 
requirement $c\sim o(1)$. It should be also noted that the vortex 
viscosities measured in YBCO are in excellent agreement\cite{futuro} with those 
measured in YBCO crystals at three different 
frequencies,\cite{tsuchiya} suggesting a common behavior for the 
electronic states in the vortex cores of SmBCO films, YBCO films and 
YBCO crystals.\\
As a final remark, we remind that the quantitative description of the 
vortex state microwave response here presented could be accomplished 
only with the essential inclusion of the enhanced QP increase in a 
magnetic field. A description in terms of fluxon motion alone would 
have given very unlikely dependences of the vortex parameters.

\section{Conclusion}

We have presented the first extensive vortex-state microwave 
characterization in SmBa$_{2}$Cu$_{3}$O$_{7-\delta}$ thin films as a 
function of the temperature and magnetic field.  The field-dependence 
of the 48 GHz complex resistivity exhibits a markedly sublinear 
behaviour.  The analysis in terms of strong pinning with weak field 
dependent superfluid depletion leads to internal contradictions, so 
that alternative interpretations have to be found.  The experimental 
results give clear indications of the irrelevance of pinning at our 
measuring frequency.  A framework including free oscillation of flux 
lines (whose resistivity is given by the flux flow expression) and an 
enhanced role of the field-induced pair breaking, due to the existence 
of lines of nodes in the superconducting gap, is able to describe the 
data qualitatively and quantitatively. The temperature dependences of 
the QP and superfluid contributions to the complex resistivity are 
fitted by the model with no free parameters.  Within this frame, we estimate 
the fluxon viscosity and we find the same values in both SmBCO 
samples.  The temperature dependence is the same as that measured in 
YBCO. Moreover, the viscosity data in SmBCO and YBCO scale together 
with a numerical factor.  Using the Bardeen-Stephen model we estimate 
the upper critical field in SmBCO and the $c$ constants, which we 
consistently find $\sim o(1)$.  We conclude that the vortex state 
resitivity in SmBCO is well described by vortex motion (that at our 
measuring frequencies takes the form of flux flow), with the 
essential contribution of the strong field dependence of the QP and 
superfluid fractional densities.  The latter is connected to the 
structure of the superconducting gap.

\section*{acknowledgements}
Useful discussions with A. Maeda and R. W\"ordenweber are warmly 
acknowledged. Support from the ESF ``VORTEX'' program is 
acknowledged. This work has been partially supported by Italian MIUR under FIRB 
``Strutture semiconduttore/superconduttore per l'elettronica 
integrata''. L.M. acknowledges financial support from MIUR under the 
same project.

%
\begin{twocolumn}

\begin{figure}
\centerline{\psfig{figure=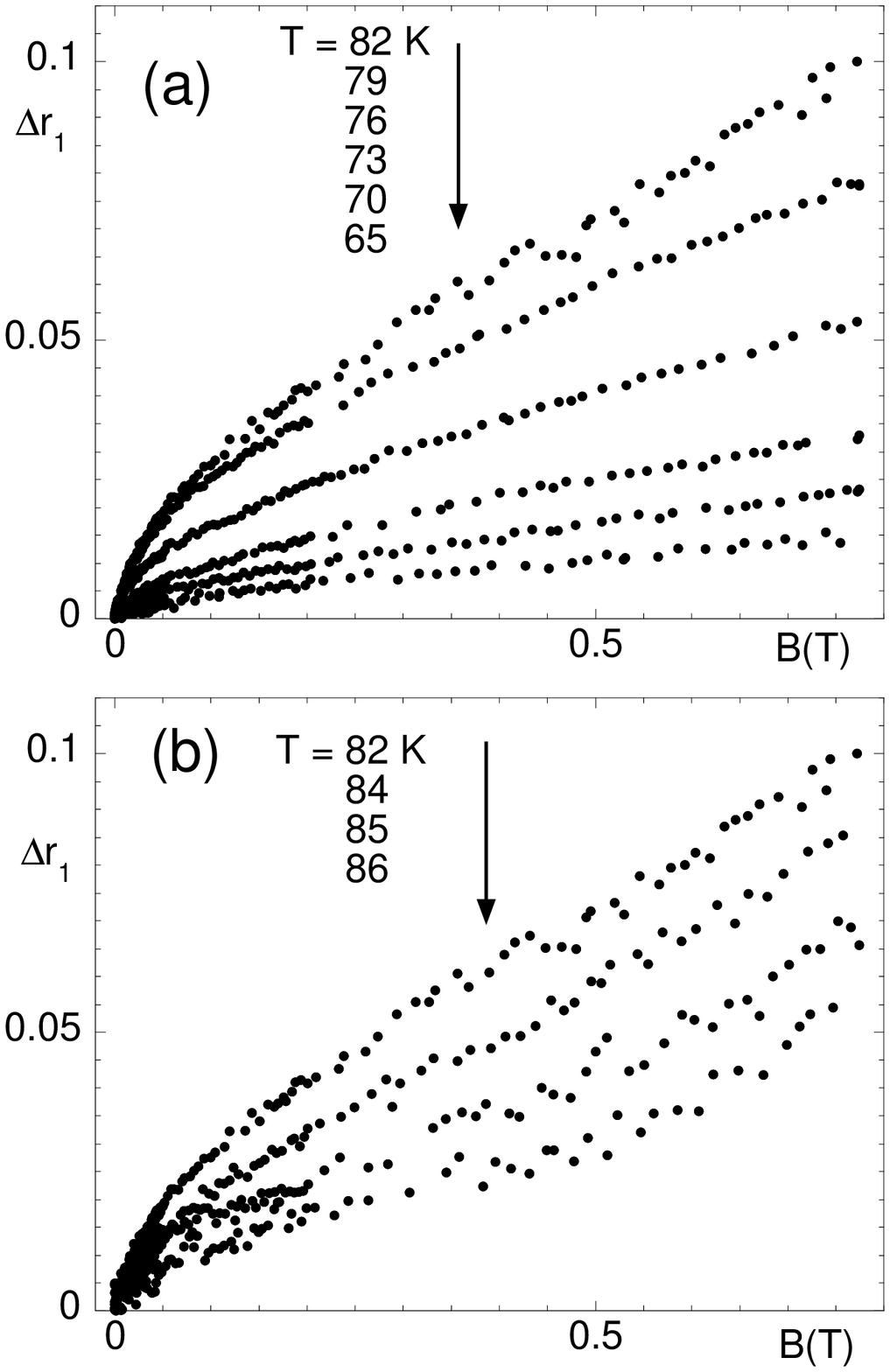,height=11.0cm,clip=,angle=0.}}
\caption{Field increase of the normalized 
real microwave resistivity $\Delta r_{1}$ in sample A at 48.2 GHz at 
selected temperatures.  Upper panel: raise of $\Delta r_{1}$ with 
increasing temperature up to $T_{max}\sim$82 K. Lower panel: the 
amplitude decreases above $T_{max}$.}
\label{r1vsB}
\end{figure}


\begin{figure}
\centerline{\psfig{figure=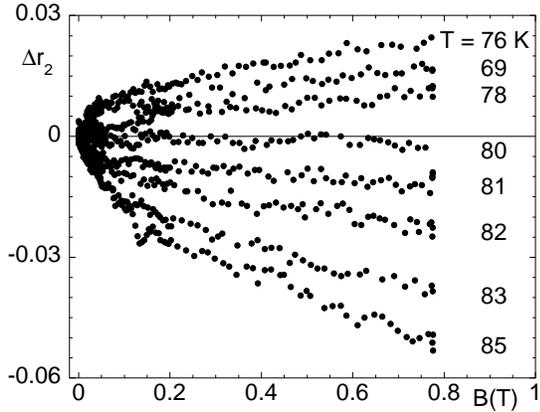,height=5.5cm,clip=,angle=0.}}
\caption{Field variation of the normalized 
imaginary microwave resistivity $\Delta r_{2}$ in sample A at 48.2 GHz 
at selected temperatures.  Up to $T_{0}\sim$80 K the field variation 
is positive, above $T_{0}$ it becomes negative.}
\label{r2vsB}
\end{figure}


\begin{figure}
\centerline{\psfig{figure=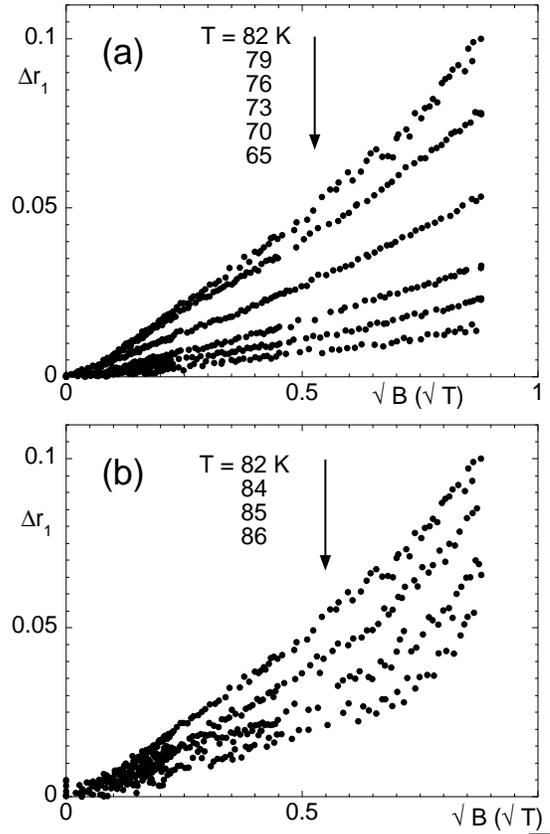,height=11.0cm,clip=,angle=0.}}
\caption{Same data as in Figure \ref{r1vsB}, plotted against 
$\sqrt{B}$.  The data show a detectable upward curvature.}
\label{r1vssqrtB}
\end{figure}


\begin{figure}
\centerline{\psfig{figure=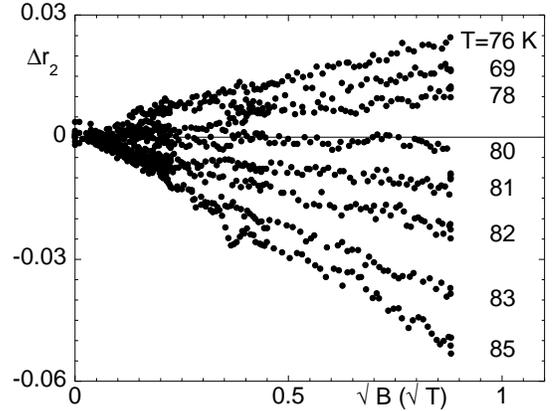,height=5.5cm,clip=,angle=0.}}
\caption{Same data as in Figure \ref{r2vsB}, plotted against 
$\sqrt{B}$.  The data exhibit a nearly perfect linear behaviour.}
\label{r2vssqrtB}
\end{figure}


\begin{figure}
\centerline{\psfig{figure=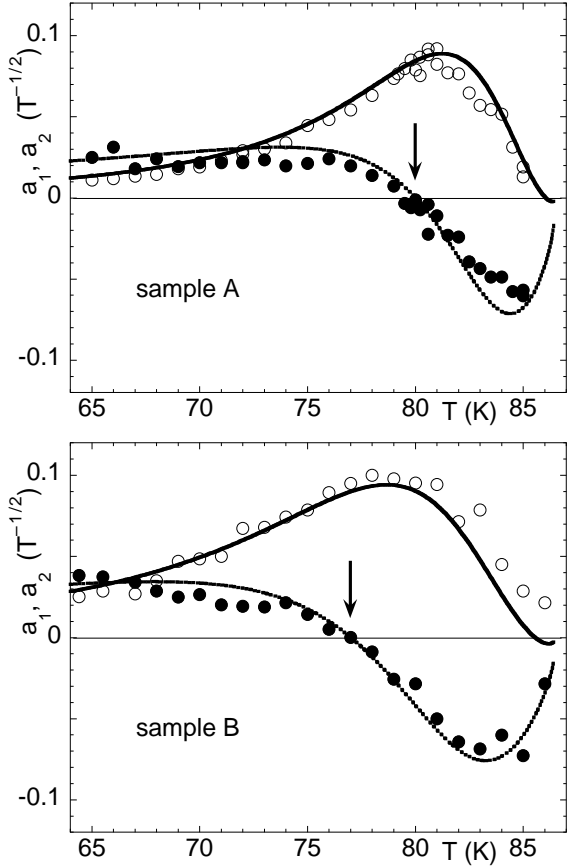,height=11.5cm,clip=,angle=0.}}
\caption{Temperature dependence of the coefficients of the $\sqrt{B}$ 
terms in the real ($a_{1}$) and imaginary ($a_{2}$) parts of the 
normalized resistivity. Open symbols: $a_{1}$, full symbols: $a_{2}$.  
Arrows mark the temperature $T_{0}$ where the imaginary resistivity is 
nearly insensitive to the applied field.  Upper panel: sample A; lower 
panel: sample B. Continuous and dashed lines are the simultaneous fits 
with Equations \ref{cc1final} and \ref{cc2final}, respectively.  For 
each pair of fits only an overall scale factor is used.}
\label{a1a2}
\end{figure}


\begin{figure}
\centerline{\psfig{figure=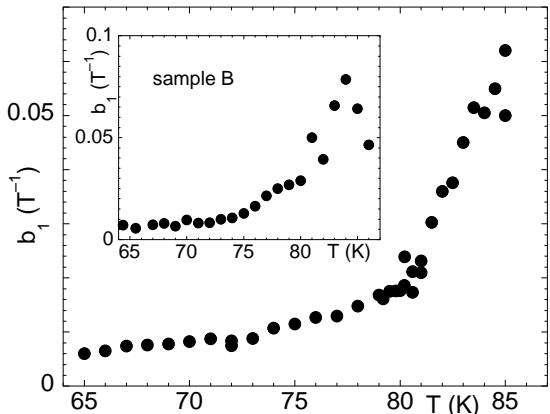,height=5.5cm,clip=,angle=0.}}
\caption{Temperature dependence of the coefficients $b_{1}$ of the 
$\sim B$ term in the real part of the normalized resistivity, related 
to the vortex contribution.  Main panel, sample A. Inset, sample B.}
\label{b1}
\end{figure}


\begin{figure}
\centerline{\psfig{figure=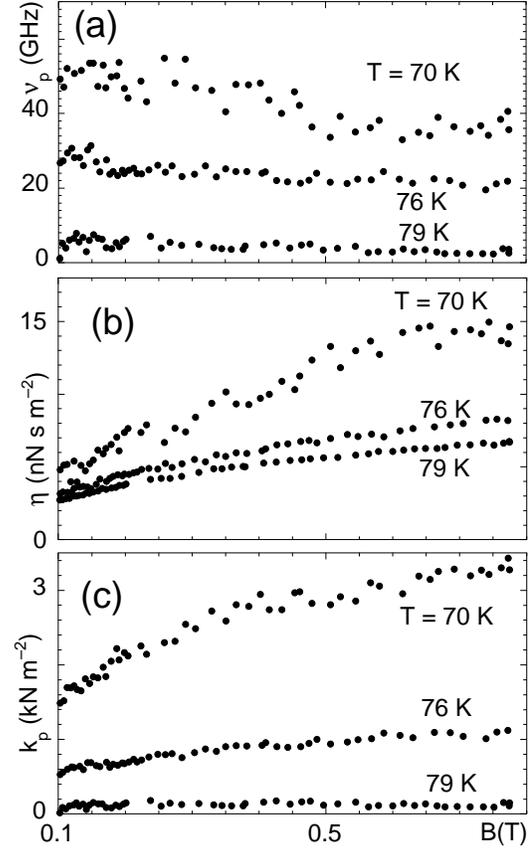,height=11.5cm,clip=,angle=0.}}
\caption{Vortex parameters as calculated from data in sample A 
accordingly to the conventional Gittleman-Rosenblum model.  (a): 
pinning frequency.  (b): vortex viscosity.  (c): pinning constant.  
The strong field dependence of the so-calculated vortex viscosity and 
pinning constant cannot be easily justified.}
\label{conventional}
\end{figure}


\begin{figure}
\centerline{\psfig{figure=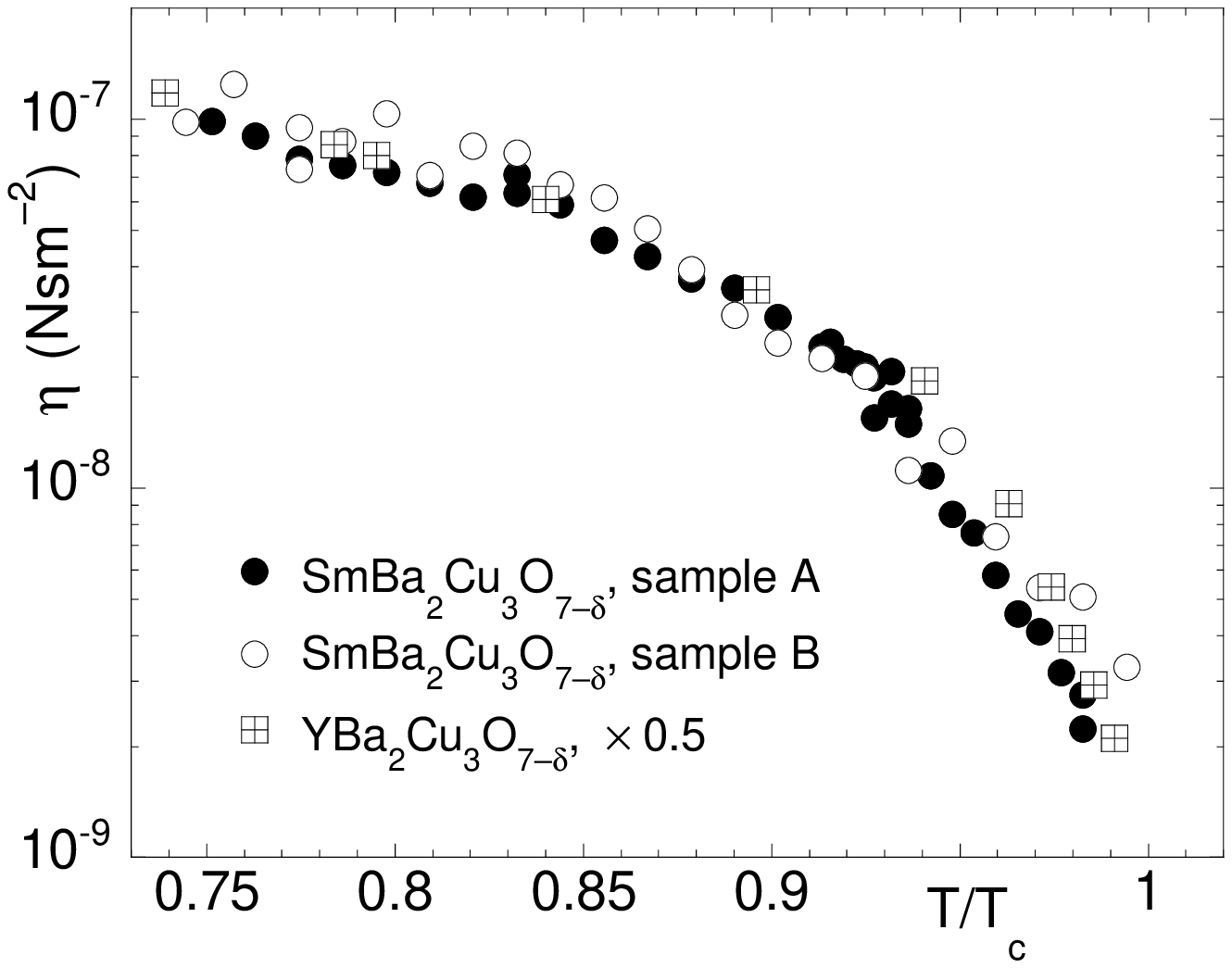,height=5.5cm,clip=,angle=0.}}
\caption{Vortex viscosity in SmBCO (full dots, sample A; open circles, 
sample B) and in YBCO (squares) as a function of the reduced 
temperature.  Data in YBCO are scaled by a factor 2 to show the 
collapse on the same curve as in SmBCO. Critical temperatures are 
$T_{c}=$86.5 K in SmBCO samples, and $T_{c}=$89.3 K in YBCO. The viscosity 
in SmBCO is the same in both samples.  The temperature dependence is 
identical in YBCO and SmBCO.}
\label{etainsieme}
\end{figure}

\end{twocolumn}
\end{multicols}


\begin{references}
\bibitem{notatau}
We have taken for simplicity a field-independent quasiparticle 
scattering time. However, it is easy to show that this choice changes 
the subsequent equations only by some numerical factor. For example, 
allowing for an electron-vortex scattering like the one proposed in 
\cite{ausloos}, one would only redefine the quantity $\frac{x_{s}}{x_{n}}$ 
in Eq.s\ref{cc1final},\ref{cc2final}.
\bibitem{notafit}
With a different choice of the temperature dependence of $x_{n0}$ and 
$x_{s0}$ one would have different $S_{0}$.  As an example, in sample A 
we have tried $x_{n0}=t^{\alpha}$ and $x_{s0}=1-x_{n0}$, with 
$\alpha=$1,2,4 (and correspondingly $S_{0}=$0.075, 0.146, 0.278 for 
our values of $T_{0}$), obtaining nearly identical fits (with the same 
scale factor).
\end{references}
\end{document}